\documentclass[twocolumn,showpacs,pre]{revtex4}

\usepackage{graphicx}
\usepackage{bm}

\newcommand{\av}[2]{ { \left\langle #2 \right\rangle_{#1} } }
\newcommand{\set}[1]{ { \{#1\} } }
\newcommand{\avx}[1]{ \av{\set{x}}{#1} }
\newcommand{\mref}[1]{(\ref{#1})}
\newcommand{\Ham}{ {\cal H} }
\newcommand{\kB}{ {k_{\rm B}} }
\newcommand{\kT}{ {\kB T} }
\newcommand{\Tr}{ \mbox{Tr} }
\newcommand{\di}{d}
\newcommand{\atanh}{ \mbox{atanh} }
\newcommand{\Min}{ {\rm min} }
\newcommand{\Max}{ {\rm max} }
\newcommand{\rhs}{\textit{rhs} }
\newcommand{\lhs}{\textit{lhs} }

\begin{document}

\title{Model fluid in a porous medium: results for a Bethe lattice}

\author{R. O. Sokolovskii}
 \altaffiliation[Also at ]{Institute for Condensed Matter Physics, 1 Svientsitskii, Lviv 79011, Ukraine}

\author{M. E. Cates}

\author{T. G. Sokolovska}
 \altaffiliation[Also at ]{Institute for Condensed Matter Physics, 1 Svientsitskii, Lviv 79011, Ukraine}%
 \email{tata@icmp.lviv.ua}

\affiliation{School of Physics, JCMB Kings Buildings, University of Edinburgh, Edinburgh EH9 3JZ, United Kingdom}

\date{\today}

\begin{abstract}   
We consider a lattice gas with quenched impurities or `quenched-annealed binary mixture' on the Bethe lattice. The quenched part represents a porous matrix in which the (annealed) lattice gas resides. This model features the 3 main factors of fluids in random porous media: wetting, randomness and confinement. The recursive character of the Bethe lattice enables an exact treatment, whose key ingredient is an integral equation yielding the one-particle effective field distribution. Our analysis shows that this distribution consists of two essentially different parts. The first one is a continuous spectrum and corresponds to the macroscopic volume accessible to the fluid, the second is discrete and comes from finite closed cavities in the porous medium. Those closed cavities are in equilibrium with the bulk fluid within the grand canonical ensemble we use, but are inaccessible in real experimental situations. Fortunately, we are able to isolate their contributions. Separation of the discrete spectrum facilitates also the numerical solution of the main equation. The numerical calculations show that the continuous spectrum becomes more and more rough as the temperature decreases, and this limits the accuracy of the solution at low temperatures.   \end{abstract} 
\maketitle

\section{\label{sec:intro}Introduction}

When fluids are adsorbed in porous materials they behave very differently from what we know in the bulk. This happens in both high-porosity materials, such as silica aerogels, and low-porosity materials, such as Vycor glass. Even an aerogel that occupies a few percent of the total volume significantly deforms and reduces the gas-liquid binodal of the fluid. At least three factors affect phase equilibrium of fluids in porous media: wetting, randomness, and confinement by the matrix. A number of models exist that emphasize one or two of those factors, and they greatly facilitate understanding of possible behaviors of the fluid in the porous media (see review in Ref.~\cite{GelbReview}). But theoretical models that take into account all three factors appear to be extremely hard to deal with. All existing microscopic theories are inconclusive even concerning the qualitative behavior of the gas-liquid binodal in the porous medium. Conventional approximation schemes yield very different results depending on the model and on the level of approximation \cite{Kierlik98,Kierlik9799}. Monte Carlo simulations suffer from long relaxation times and are performed in a very limited simulation box \cite{Page96,Alvarez99,GelbReview}. This makes any exactly treatable model especially desirable. We shall consider such a model, which features all the three factors. It derives from the well known lattice gas model of a fluid. We shall consider here only the simplest representation of a random porous medium (quenched impurities of equal size). The model is exactly solvable on the Bethe lattice, in the sense that we can derive a closed integral equation for the local field distribution. Related approaches, leading to similar equations, were previously known in the context of spin glasses \cite{Katsura79,Mezard00} and the Random Field Ising model \cite{Bruinsma84,NumSolution}. The resulting integral equation has an interesting structure, which means that although numerical solution is relatively straightforward at high temperature, as temperature is lowered the field distribution (and hence also the numerics) becomes more and more involved. This problem is also worse close to the percolation threshold of the porous medium. 

The Bethe lattice is defined on a large uniformly branching tree of which the Bethe lattice is the part far from any perimeter site. (This is distinct from a Cayley tree which includes all parts.) The Bethe lattice should not be confused with the Bethe approximation, which usually denotes a set of mean-field-like self-consistency equations for order parameters. The Bethe approximation is closely related to the cluster variation method or cluster approximation which derive these equations from truncated cluster expansion of either the free energy or entropy functional employing several variational parameters. For many models these approximations become exact on the Bethe lattice, but not in the case studied here (except for special choices of parameters). This is a part of the motivation for the exact analysis presented below. Another important goal is to rectify a shortcoming of the grand canonical ensemble when dealing with fluids in porous media. This ensemble allows fluid to equilibrate throughout the pore space, including closed pores that are inaccessible to fluid in reality. For the Bethe lattice we are able to isolate and remove the effects of these closed pores.  

We shall formulate the model and present the basic equations in Section~\ref{sec:model}. The analysis of Section~\ref{sec:sol} will bring forward the notion of finite clusters and their importance for a correct description of the fluid. Details of the numerical algorithm and computer-aided results will be presented in Section~\ref{sec:numeric}. Section~\ref{sec:conclusion} contains our conclusions.

\section{\label{sec:model}The model}

In lattice models of a fluid, the particles are allowed to occupy only those spatial positions which belong to sites of a chosen lattice. The configurational integral of a simple fluid is thereby replaced by the partition function   
\begin{eqnarray}
&&Z= \Tr \exp \left( -\beta\Ham  \right),~\beta=1/(\kT)
,\nonumber\\
&&\Ham=H-\mu N= -\frac12\sum_{ij}I_{ij}n_i n_j-\mu\sum_i n_i
,
\label{simpleLG}
\end{eqnarray}
where $n_i$, which equals 0 or 1, is the number of particles at site $i$ ($i=1\cdots V$, where $V$ is the number of the lattice sites \cite{footnote:limit}). $\Tr$ means a summation over all occupation patterns. The total number $N$ of particles is allowed to fluctuate; $\mu$ is a chemical potential, which should be determined from the relation  
\begin{equation}
N=\av{\Ham}{\sum_i n_i};~
\av{\Ham}{\cdots}=Z^{-1}\Tr(\cdots)\exp(-\beta\Ham)
.
\end{equation}
Since particles cannot approach closer than the lattice spacing allows, lattice models automatically preserve one essential feature of the molecular interaction: non-overlapping of particles. The lattice fluid with nearest-neighbor attraction is known to demonstrate the gas-liquid transition only. Nevertheless, the lattice gas with interacting further neighbors possesses a realistic (which means argon-like) phase diagram with all the transitions between the gaseous, liquid, and solid phases being present \cite{HallStell}. In this paper we deal only with fluid phases and restrict ourselves to the nearest neighbor interaction.  

The fluid adsorbed in the porous solid can be thought of as residing on a lattice, of which a fraction of sites are excluded or pre-occupied by particles of another sort. Although in practice these blocked sites, representing the solid, must form a connected network (if the solid is to remain static) we ignore this here and allow sites to be blocked at random. Thus, we have to consider the lattice gas with quenched impurities, whose Hamiltonian is given by
\cite{Kierlik98}
\begin{equation}
\Ham=-I\sum_\av{}{ij} n_i n_j  - K\sum_\av{}{ij} n_i x_j 
-\mu \sum_i n_i
,
\label{Ham}
\end{equation}
where $\sum_\av{}{ij}$ means summation over all nearest neighbor sites, and quenched variables $x_i$ describe the presence ($x_i=1$) or absence ($x_i=0$) of a solid particle at site $i$. Each site of the lattice can be either empty ($x_i=0$, $n_i=0$), occupied by a fluid particle  ($x_i=0$, $n_i=1$), or filled with a particle belonging to the porous solid and therefore inaccessible to the fluid ($x_i=1$, $n_i=0$).  The first term of Eq.~\mref{Ham} describes the nearest neighbor attraction between the fluid particles ($I>0$), the second one corresponds to the fluid-solid attraction ($K>0$) or repulsion ($K<0$) on the nearest neighbor sites. The hard-core repulsion is taken into account by mutual exclusiveness of particles: $x_i+n_i\le1$.

It is quite customary to establish the connection between lattice gases and the Ising model. Indeed, a simple substitution $S_i=2n_i-1$ transforms \mref{simpleLG} into the Ising Hamiltonian. In the Ising model language $S_i=-1$ means an empty site, and $S_i=1$ corresponds to a site occupied by a fluid particle. The current model \mref{Ham} is equivalent to the diluted Ising model with random surface field $\Delta$:
\begin{eqnarray}
&&\Ham=-\sum_\av{}{ij}J_{ij} S_i S_j  -\sum_i h_i S_i +const
,\label{IsHam}
\\
&&J_{ij}= J y_i y_j;\ h_i=y_iH+\Delta_i;\ \Delta_i=y_i \bar H \sum_jx_j
,\label{IsHam2} 
\\
&&J=I/4;\ H=\mu/2+zI/4;\ \bar H=K/2 - I/4
.\label{IsHam3} 
\end{eqnarray}
Here $y_i=1-x_i$ describes the accessibility of site $i$ to the fluid particles; the constant term in \mref{IsHam} embraces several pieces that depend only on quenched variables and therefore do not contribute to the thermodynamics; $z$ is the coordination number of the lattice; the sum on $j$ in \mref{IsHam2} spans nearest neighbor sites to site $i$. When deriving \mref{IsHam} we used the equality $n_i=n_iy_i$. $\Delta_i$ is a random field correlated to surface sites: $\Delta_i\ne0$ at the sites that are accessible to the fluid ($y_i=1$), and are in contact with the solid (at least one of the neighbor sites has $x_j\ne0$). When $\bar H=0$, Eq.~\mref{IsHam} is a Hamiltonian of the diluted Ising ferromagnet, and we shall frequently refer to this limiting case throughout the paper.

The grand thermodynamic potential of the model is 
\begin{eqnarray}
&&F=-\kT \avx{\ln \sum_{S_1}\cdots\sum_{S_V} \exp(-\beta\Ham)}
,\label{F}
\\
&&\avx{ (\cdots) }=\sum_{x_1=0}^1\cdots\sum_{x_V=0}^1\rho(\set{x_i}) (\cdots)
,\label{avx}
\end{eqnarray}
where the distribution function $\rho(\set{x_i})$ is a quenched one that describes the distribution of solid particles in space.

\begin{figure}
\includegraphics[width=6cm]{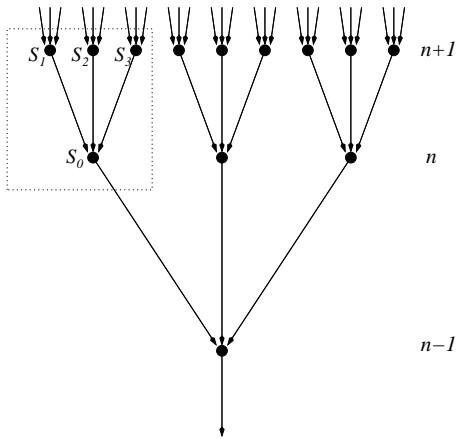}
\caption{\label{fig:lattice}A schematic picture of the Bethe lattice with coordination number $z=4$. The arrowheads do not mean that the links are asymmetric, they just illustrate the way the partition function calculation proceeds (see text).}
\end{figure}

We shall study this model on the Bethe lattice, which permits us to calculate an exact thermodynamic potential. It is known that many problems on tree-like structures, such as that depicted in Fig.~\ref{fig:lattice}, can be solved iteratively. The term `Bethe lattice' refers to an infinitely deep part of such a tree.

Each site on the $n$th level has $k=z-1$ neighbors at the ($n+1$)th level and one neighbor at the ($n-1$)th level. The partition function of the model can be calculated recursively, from upper levels to the bottom. For example, let us consider the cluster of sites framed by the dashed box in Fig.~\ref{fig:lattice}, where $k = 3$. The following relation (which holds for all $k$) shows how tracing out spins at the upper level forms an `effective field' on the lower level:
\begin{eqnarray}
&&	\sum_{S_1}\cdots\sum_{S_k} 
\exp(\beta S_0\sum_{l=1}^k J_{0l}S_l+\beta \sum_{l=0}^k h_lS_l)=
\nonumber \\
&&=
[\prod_{l=1}^k C(J_{0l},h_l)]\exp(\beta \tilde h_0S_0)
,\\
&&
\tilde h_0=h_0 + \sum_{l=1}^k U(J_{0l},h_l)
,\\
&&
U(J,h)=\beta^{-1}\atanh[\tanh(\beta J)\tanh(\beta h)]
,\\
&&
C(J,h)=2\frac{\cosh(\beta J)\cosh(\beta h)}{\cosh[\beta U(J,h)]}.
\end{eqnarray}
We have thus obtained an effective field $\tilde h_0$ which is a sum of the original field $h_0$ and $k$ contributions from the upper branches. We can proceed this way recursively downward, giving  
\begin{equation}
\tilde h_i=h_i + \sum_{j=1}^k U(J_{ij},\tilde h_j)
\label{h0}
\end{equation}
at each level. In the non-random case, when $h_i=h$ and $J_{ij}=J$, the effective field deep inside the tree satisfies the stable point relation that follows from Eq.~\mref{h0}:
\begin{equation}
\tilde h=h + k U(J,\tilde h)
.\label{stable}
\end{equation}
For our model \mref{IsHam}, we can expect that a properly defined effective field \textit{distribution} tends to a stable limiting form at deep levels of the tree. 

Let us consider the probability $Q_i(y,h)\di h$ that $y_i=y$ and $\tilde h_i=h$:
\begin{equation}
Q_i(y,h)=\avx{\delta_{yy_i}\delta(h-\tilde h_i)}
,
\end{equation}
where $\delta(h)$ stands for the Dirac distribution, and $\delta_{ab}$ is a Kronecker's delta ($\delta_{ab}$ equals 1 if $a=b$, and equals zero otherwise). $Q_i(0,h)$ is the effective field distribution at a matrix site, and $Q_i(1,h)$ is the field distribution for a site accessible to fluid. It is easy to see from \mref{h0} and \mref{IsHam2} that when $y_i=0$, $J_{ij}=0$ and $h_i=0$, and the effective field $\tilde h_i$ equals zero, thus $Q_i(0,h)$ does not depend on $i$ and has
a simple form
\begin{equation}
 Q_i(0,h)=Q(0,h)=c_0\delta(h)
\end{equation}
where we define
\begin{equation}
c_y\equiv\avx{\delta_{yy_i}},
\end{equation}
so that $c_0$ is a fraction of the sites occupied by the matrix. When we move recursively from the surface of the tree deeper into its interior, $Q_i(1,h)$ changes, but should tend to a fixed point at infinitely deep levels of the tree: $Q_i(1,h)\to Q(1,h)$. It follows from \mref{h0} that the resulting distribution obeys
\begin{widetext}
\begin{equation}
Q(1,h)=
\avx{y_i
    \int[\prod_{j=1}^k \di h_j Q(y_j,h_j)]
    \delta(h-h_i-\sum_{j=1}^k U(J_{ij},h_j))
}
.\label{recursion}
\end{equation}
\end{widetext}

In writing this equation, and throughout the following, we now specialize to the case where there is no correlation of the $\set{y_i}$ variables (which specify the sites occupied by the solid matrix) beyond nearest neighbor correlations, which are permitted. In this case, taking into account that $k$ upper branches do not interact until they meet the chosen site, the fields $\set{h_j}$ are uncorrelated. (In more general cases, the $y_j$ at different branches are correlated; the joint distribution $Q(y_1,h_1,\cdots,y_k,h_k)$ cannot be decoupled as a product of $Q(y_j,h_j)$; and \mref{recursion} is invalid.) Using the integral representation of the delta function
\begin{equation}
    \delta(h)=\int \frac{\di\xi}{2\pi}\exp(i\xi h)
\label{delta}
\end{equation}
we can factorize the delta function in \mref{recursion} and get
\begin{eqnarray}
\label{recursion2}
Q(1,h)&=&
\Big\langle
    \int\frac{\di\xi}{2\pi}y_i\exp(i\xi(h-h_i))
\times  \\ \nonumber
&\times&    \prod_{j=1}^k \int\di h_j Q(y_j,h_j)
    \exp(-i\xi U(J_{ij},h_j))
\Big\rangle_\set{x}. 
\end{eqnarray}
Configurational averaging with respect to the quenched variables $\avx{\cdots}$ in \mref{recursion2} can now be performed explicitly, and it yields
\begin{eqnarray}
Q(h)&=&\int \frac{\di\xi}{2\pi}\exp(i\xi(h-H))\tilde Q^k(\xi)
,\label{main}\\
\tilde Q(\xi)&=&p_0\exp(-i\xi\bar H)+p_1\int\di h' Q(h')\exp(-i\xi u(h'))
,\nonumber
\end{eqnarray}
where we introduced the notations $Q(h)=Q(1,h)/c_1$; $u(h)=U(J,h)$; $p_a=w_{a1}/c_1$; and 
\begin{equation}
w_{ab}\equiv\avx{\delta_{ay_i}\delta_{by_j}}
\end{equation} 
is the probability to find $y_i=a$ and $y_j=b$ at a randomly chosen pair of nearest neighbor sites $i$ and $j$. For example, $w_{11}=\avx{y_iy_j}$ is a probability that both sites in the pair are accessible to fluid. For each site accessible to the fluid, $p_1$ specifies the probability that a given nearest neighbor site is accessible as well; $p_0=1-p_1$ is, correspondingly, the probability that this neighbor site is occupied by the solid.

Note that $Q(h)$ specifies the distribution of the effective field created by $k$ `upper branches'. The complete one-site field deep inside the tree consists of contributions coming from \textit{all} $z$ nearest neighbors, and its distribution takes into account $z$ equivalent branches
\begin{eqnarray}
Q_z(1,h)&=&\avx{y_i\delta( h-h_i - \sum_{j=1}^z U(J_{ij},\tilde h_j) )}
\nonumber \\ 
&=&c_1\int \frac{\di\xi}{2\pi}\exp(i\xi(h-H))\tilde Q^z(\xi)
,\label{Qz(h)}\\
Q_z(0,h)&=&c_0\delta(h)
.
\end{eqnarray}
This distribution determines the one-site average values. For example, in the magnetic interpretation of the model \mref{Ham}, this total one-site field permits to calculate the average magnetization (per magnetic site)
\begin{equation}
m=\av{x}{\av{\Ham}{n_iS_i}}/c_1=\int\di h Q_z(h)\tanh(\beta h)
.
\end{equation}
Calculation of other thermodynamic properties is less straitforward. The thermodynamic potential of the tree is a sum of `site' and `link' contributions \cite{Katsura79,Mezard00}:
\begin{equation}
\beta F(\set{x})=k\sum_i\ln\Tr_i\exp(-\beta \Ham_i)
    -\sum_\av{}{ij}\ln\Tr_{ij}\exp(-\beta \Ham_{ij})
,\label{F2}
\end{equation}
where $\Ham_i$ and $\Ham_{ij}$ are Hamiltonians of one site and a pair of sites, respectively, and the traces act only on the corresponding local degrees of freedom. The configurationally averaged thermodynamic potential of the Bethe latice must take into account only deep levels of the tree. We do this by assigning the stable point field distribution to all the sites when doing configurational averaging. The Bethe lattice thermodynamic potential $F$ is, therefore, given by
\begin{widetext}
\begin{eqnarray}
\beta F/N&=&\beta \avx{ F(\set{x}) }/N
\label{Fav} 
\\ \nonumber
     &=&k\sum_{y_i=0}^1\int\di h_i Q_z(y_i,h_i)\ln\Tr_i\exp(-\beta \Ham_i)
    -\frac{z}{2}\sum_{y_i=0}^1\sum_{y_j=0}^1 \frac{w_{y_iy_j}}{c_{y_i}c_{y_j}}\int\di h_i Q(y_i,h_i)\int\di h_j
Q(y_j,h_j)\ln\Tr_{ij}\exp(-\beta \Ham_{ij})
,
\end{eqnarray}
\end{widetext}
where $\Ham_i$ and $\Ham_{ij}$ depend on both fluid ($S$) and matrix ($y$ or, equivalently, $x$) variables:
\begin{equation}
\Ham_i=h_i S_i;\;\Ham_{ij}=h_i S_i+h_j S_j+J_{ij}S_iS_j
.
\end{equation}

\section{\label{sec:sol}The solution}

The result of the previous section is that thermodynamics of the model is determined by the one-site field distribution $Q(h)$, and this distribution satisfies integral equation \mref{main}. Equations of this type have been known previously in the context of spin glasses \cite{Katsura79,Mezard00} and the Random Field Ising model \cite{Bruinsma84,NumSolution}. We did not encounter this kind of equation for the current model in the literature, even for the limiting case of diluted ferromagnet ($\bar H=0$). In this section we investigate the ways of solving this equation.

For the diluted Ising ferromagnet in zero external field, when $H=\bar H=0$, $Q(h)=\delta(h)$ is always a solution of Eq.~\mref{main}, and it yields zero magnetization ($\av{}{S_i}=0$) or, in the fluid model language, the occupancy number equals one half ($\av{}{n_i}=1/2$) at all sites. This is known as the trivial solution in all calculations leading to an analytical result, and this is the only solution at the temperatures above critical. What are the nontrivial solutions? 

When there is no solid medium ($c_1=1$) or when the solid sites are arranged in a nonporous block or slab with a smooth surface ($p_1=1$), the solution is $Q(h)=\delta(h-\tilde h)$, and $\tilde h$ is given by Eq.~\mref{stable}. This equation has a nontrivial solution ($\tilde h\ne0$) at low temperatures: $T<T_c$, $T_c=J/[\kB\atanh(1/k)]$. In this case the results of the cluster (or Bethe) approximation become exact. 

When quenched chaos is present ($p_1\ne1$), and $H=\bar H=0$, the trivial solution becomes unstable at temperatures below 
\begin{equation}
T_c=J/[\kB\atanh(1/(kp_1))]
\label{Tc}
\end{equation}
(this is an accurate result), signifying the appearance of the spontaneous magnetic order ($\av{}{S_i}\ne0$).
Importantly, though, the cluster approximation is not accurate when both magnetic ordering and quenched disorder are present. In the general case ($\bar H\ne0$, $H\ne0$) we do not have any simple formula for the critical temperature or the field distribution. In what follows, we study the structure of this distribution more closely. 

First we shall consider the simple limiting case of diluted magnet without surface field ($\bar H=0$) \cite{footnote:case}. In this case matrix sites break the links between spins, but do not introduce any field that breaks the spin flip symmetry. It follows from \mref{Tc} that when $p_1<1/k$, $T_c$ does not exist, and the spontaneous magnetic order does not appear. (The reason is that in this case the system is not percolated: there is no infinite cluster of linked spins, only finite clusters, and in finite systems the spin flip symmetry cannot be spontaneously broken.) Let us realize that in this system for any matrix (at any value of $p_1$, except 0 or 1) there is a finite fraction \cite{footnote:precise} of spins whose links are all broken, because all their neighboring sites are occupied by the matrix. The effective field at those sites equals just an external field $H$. Thus it is likely that the solution $Q(h)$ contains a term in $\delta(h-H)$. In the case of zero external field ($H=0$) one easily finds that a solution of the form $Q(h)=A_0\delta(h)+q(h)$ does satisfy Eq. \mref{main}, and one gets 
\begin{eqnarray}
	A_0&=&(p_0+p_1A_0)^k,
\label{A0}\\
	q(h)&=&\int\frac{\di\xi}{2\pi}\exp(i\xi h)
\times\nonumber\\&&\times
[\{p_0+p_1A_0 + \tilde q(\xi)\}^k - A_0],
\\
  \tilde q(\xi)&=&p_1\int\di h' q(h')\exp(-i\xi u(h'))
\nonumber,
\end{eqnarray}
where we expect $q(h)$ to be a nonsingular function, because a delta function positioned in any other place ($A_x\delta(h-a)$, $a\ne0$) or a set of such delta functions does not lead to selfconsistent equations of the required form. 

Let us show that $A_0$ has a simple physical meaning, namely, it equals the probability that all $k$ upper branches are finite. Consider first the probability $P_f$ that a branch is finite. It consists of two possibilities: either the first link is broken (with probability $p_0$), or it is unbroken, but all farther branches connected to it are finite (with probability $p_1P_f^k$). This yields $P_f=p_0+p_1P_f^k$. Then, the probability that $k$ given branches are finite equals $A_0=P_f^k$ and satisfies Eq.~\mref{A0}. Obviously, Eq.~\mref{A0} always has the solution $A_0=1$, which leads to the trivial solution for $Q(h)$ because of the normalization condition $\int Q(h) \di h=1$. $A_0=1$ is the only solution when $p_1<1/k$, as discussed above. At the percolation point ($p_1=1/k$) it is a two-fold solution, which signifies a bifurcation. When $p_1>1/k$, a solution for $A_0$ in the interval (0,1) always exists.

When $H\ne0$, the previous ansatz that $Q(h)=A\delta(h-H)+q(h)$ does not work: $\delta(h-H)$ placed into the \rhs of Eq. \mref{main} generates, under iteration of the equation, a set of $k$ different delta functions, as does any delta function positioned in some other place. This seemingly rules out the idea of finding the discrete levels in this case, but a fully solvable case $z=2$ (the linear chain: see Appendix \ref{app:A}) resurrects the hope. In that case solution contains an \textit{infinite} series of delta functions. Therefore, in the general case we shall seek the solution in a form of a sum of an infinite set of delta functions and a nonsingular function:
\begin{equation}
Q(h)=\sum_{l=1}^\infty a_l\delta(h-h_l)+q(h)
\label{eqN}.  
\end{equation}
We now show that such a solution can really be obtained. The point is that, similarly to the zero field case ($H=0$), the equation for the discrete spectrum separates from the equation for the non-singular part $q(h)$. Placing \mref{eqN} into Eq.~\mref{main}, we get the equation for the discrete spectrum
\begin{eqnarray}
\sum_{l=1}^\infty a_l\delta(h-h_l)&=&\int \frac{\di\xi}{2\pi}\exp(i\xi(h-H))
\label{discr1}
\times\\&&\nonumber
[p_0+p_1\sum_{l=1}^\infty a_l\exp(-i\xi u(h_l))]^k
.
\end{eqnarray}
The number of delta functions is infinite, but their cumulative weight remains finite, and 
\begin{equation}
\sum_{l=1}^\infty a_l=A_0
. \label{weight}
\end{equation}
The latter relation can be obtained by integrating Eq.~\mref{discr1} with respect to $h$ and noting that the total weight of delta functions satisfies Eq.~\mref{A0}. Eq.~\mref{weight} means that the obtained discrete spectrum is just the single $h=0$ level that we had in the case $H=0$ split by the external field. Each $(a_l,h_l)$ pair corresponds to a certain finite tree `growing' up from the given site. For example, $(a_1,h_1)=(p_0^k,H)$ corresponds to the `zero' tree, when all $k$ links to the upper sites are broken; naturally $h_1=H$, and $a_1$ equals the probability that all the neighboring $k$ sites at the upper level are inaccessible.

Now let us turn back to the most general case, when both the surface and the external fields are present ($\bar H\ne0$, $H\ne0$). Inserting into \mref{main}
\begin{equation}
Q(h)=s(h)+q(h),\;s(h)\equiv\sum_{l=1}^\infty a_l\delta(h-h_l)
\label{Q1}
\end{equation}
one can obtain explicit expressions for the discrete $s(h)$ and continuous $q(h)$ parts of the spectrum:
\begin{eqnarray}
s(h)
&=&\int \frac{\di\xi}{2\pi}\exp(i\xi(h-H)) \tilde s^k(\xi)
,\label{discr}\\ 
\label{contin}
q(h)&=&\int \frac{\di\xi}{2\pi}\exp(i\xi(h-H))
\times\\&&\times\nonumber
[\{\tilde q(\xi)+\tilde s(\xi)\}^k - \tilde s^k(\xi)]
,
\end{eqnarray}
where 
\begin{eqnarray}
\label{addedlabel}
\tilde s(\xi)&=&p_0\exp(-i\xi\bar H)
+p_1\sum_{l=1}^\infty a_l\exp(-i\xi u(h_l))
,\\ 
\tilde q(\xi)&=&p_1\int\di h' q(h')\exp(-i\xi u(h'))
\label{addedlabel2}
.
\end{eqnarray}

Note that Eq.~\mref{addedlabel} for the discrete spectrum has a closed form and does not depend on $q(h)$, whereas the equation for $q(h)$ does include $s(h)$. This is because the discrete spectrum comes from finite trees, whereas the continuous spectrum is an attribute of the infinite volume. Finite branches know nothing about the infinite volume, therefore $s(h)$ does not depend on $q(h)$. At the same time, at any given site finite branches may connect onto the infinite tree; therefore the equation for $q(h)$ does depend on the discrete spectrum $s(h)$. 

Let us consider the structure of Eq.~\mref{contin}. It can be rewritten in the form
\begin{equation}
q(h)=\int \frac{\di\xi}{2\pi}\exp(i\xi(h-H))
\sum_{l=1}^k\pmatrix{k\cr l}\tilde q^l(\xi)\tilde s^{k-l}(\xi)
\label{contin2}
.
\end{equation}
Let us recall that by construction of equation \mref{main} for $Q(h)$ if we insert some field distribution as $Q(h)$ in the \rhs, we obtain in the \lhs the field distribution one level deeper the tree. The same property holds for Eq.~\mref{contin2}. Then Eq.~\mref{contin2} permits a clear interpretation: each term of $\sum_l$ mixes into the next-level distribution the possibility that $l$ branches are infinite, with the other $k-l$ finite; $l$ runs from 1 to $k$, the $l=0$ term was rightly eliminated by the last term in Eq.~\mref{contin}, because this term would give rise to the possibility of a finite tree and would give a discrete spectrum contribution. This interpretation makes $\tilde s(\xi)$ and $\tilde q(\xi)$ responsible for finite and infinite branches respectively. The surface field $\bar H$ does not enter Eq.~\mref{contin} explicitly; its influence on $q(h)$ is mediated by the discrete spectrum $s(h)$. In a way, $s(h)$ (or, equivalently, $\tilde s(\xi)$ \cite{footnote:correspondence}) encapsulates the surface influence. The only explicit reminder of a porous medium in the equation for the continuous spectrum is the $p_1$ multiplier in Eq.~\mref{addedlabel2}, indicating that links between sites are not always present.

The complete effective field acting on an accessible site $Q_z(1,h)$ (defined by \mref{Qz(h)}) likewise consists of a discrete spectrum $s_z(h)$, and a continuous one $q_z(h)$:
\begin{eqnarray}
Q_z(1,h)&=&c_1s_z(h)+c_1q_z(h)
\label{P1} \\ 
s_z(h)&=&\int \frac{\di\xi}{2\pi}\exp(i\xi(h-H))\tilde s^z(\xi)
,\label{Pcont}\\
q_z(h)&=&\int \frac{\di\xi}{2\pi}\exp(i\xi(h-H))
\times\\&&\times\nonumber
[\{\tilde q(\xi)+\tilde s(\xi)\}^z-\tilde s^z(\xi)]
.
\end{eqnarray}
The total weight of the discrete part
\begin{equation}
x=\int s_z(h) \di h=A_0^{z/k}
\label{xx}
\end{equation}
equals the probability that all $z$ branches connected to this site are finite. When this happens, the site belongs to a finite closed volume. 

When considering the fluid in the grand canonical ensemble (as we have done so far) all sites are in equilibrium with an implicit bulk fluid, whose temperature and chemical potential are parameters that describe this equilibrium. But in real experiments fluid particles cannot access any disconnected pore volumes (described by $s_z(h)$), and those parts of the system do not respond to changes of the chemical potential. Were we considering a magnetic system, spins in those finite separated clusters would be able to react to changes of external field $H$, and thermodynamic potential \mref{Fav} would be valid. In contrast, in order to describe the experimental situation for a fluid in porous medium, we have to exclude the contribution of finite volumes to the thermodynamic potential. Fortunately, we are now able to separate this contribution because finite branches always generate a discrete spectrum, in contrast to the infinite cluster, which yields the continuous distribution. 

Inserting \mref{Q1} and \mref{P1} into the expression for the thermodynamic potential \mref{Fav}, we can identify three distinct contributions of finite volumes
\begin{eqnarray}
\label{excluded} 
\nonumber&&
kc_1\int\di h_i s_z(h_i)\ln\Tr_i\exp(-\beta \Ham_i)\Big|_{y_i=1}
,\\ 
\nonumber && 
    -\frac{z}{2}w_{11}\int\di h_i s(h_i)\int\di h_j s(h_j)\ln\Tr_{ij}\exp(-\beta \Ham_{ij})\Big|_{y_i=1,y_j=1}
,\\ 
\nonumber && 
    -zw_{10}\int\di h_i s(h_i)\ln\Tr_{ij}\exp(-\beta \Ham_{ij})\Big|_{y_i=1,y_j=0}
.
\end{eqnarray}
The first term above corresponds to an accessible site whose branches are all finite, the second one describes a linked pair of accessible sites with outer branches finite, and in the third term one site of the pair is accessible to the  fluid, but all its branches are finite. Note that the division into these three terms does not reflect the presence of different types of site but reflects the subdivision into site and link contributions in \mref{F2}. 

Dropping these `finite volume' contributions (and constant terms corresponding to the sites occupied by the matrix) we obtain the thermodynamic potential of the fluid in the accessible volume as:
\begin{widetext}
\begin{eqnarray}
\label{Ffluid} 
\beta F/N&=&
    kc_1\int\di h_i q_z(h_i)\ln\Tr_i\exp(-\beta \Ham_i)\Big|_{y_i=1}
\nonumber \\ && 
    -\frac{z}2 w_{11}\int\di h_i q(h_i)\int\di h_j [q(h_j)+2s(h_j)]\ln\Tr_{ij}\exp(-\beta \Ham_{ij})\Big|_{y_i=1,y_j=1}
\nonumber \\ && 
    -zw_{10}\int\di h_i q(h_i)\ln\Tr_{ij}\exp(-\beta \Ham_{ij})\Big|_{y_i=1,y_j=0}
.
\end{eqnarray}
\end{widetext}

The procedure of `contribution classification' is especially simple when we consider the expression for the density (per free volume) $n=\avx{\av{\Ham}{n_i}}/c_1$. The grand canonical ensemble (GCE) predicts
\begin{equation}
n=\int\di h [q_z(h)+s_z(h)] \cdot \frac12[1+\tanh(\beta h)]=\frac{m+1}2
\label{nGC}
,
\end{equation}
whereas when we take into account that closed finite pores are inaccessible to the fluid the microscopic fluid density at those sites has to be set to zero, and we obtain the corrected, `infinite cluster' (IC) expression
\begin{eqnarray}
n &=&	\int\di h \{ q_z(h) \cdot \frac12[1+\tanh(\beta h)]+s_z(h) \cdot 0 \}
\nonumber\\
  &=&	[
  	1-x+\int\di h q_z(h) \tanh(\beta h)
  ]/2
\label{nIC} 
\\ \nonumber
  &=& \frac{m+1}2 - \Delta n
,
\end{eqnarray}
where $x$ is a fraction of free sites belonging to the finite cavities, defined by Eq.~\mref{xx}, and the difference between the GCE and IC densities 

\begin{equation}
\Delta n=\int\di h s_z(h) \cdot \frac12[1+\tanh(\beta h)]
\label{dn}
\end{equation}
is simply the number of fluid particles in finite cavities per number of the free sites.

\section{\label{sec:numeric}Numerical calculations}

In order to solve the main equation \mref{main} we have to calculate the discrete spectrum \mref{discr} and use it when solving for the continuous part of the field distribition via \mref{contin}. The procedure of the discrete spectrum calculation is iterative. We place the zeroth approximation $s(h)=p_0^k\delta(h-H-k\bar H)$ into the \rhs of Eq.~\mref{discr} and obtain in the \lhs a sum of $k+1$ delta functions, one of which coincides with the initial one and has the correct weight $p_0^k$. Placing the new approximation into the \rhs of Eq.~\mref{discr} recursively, we can obtain as many terms as we need. The unpleasant feature is that the number of levels in the spectrum grows exponentially during the iterations, and many levels have negligible weight. Fortunately, one can drop the low-weight levels at each step, because they produce only lower-weight levels during the following iterations. 

\begin{figure}
\includegraphics[angle=-90,width=\columnwidth]{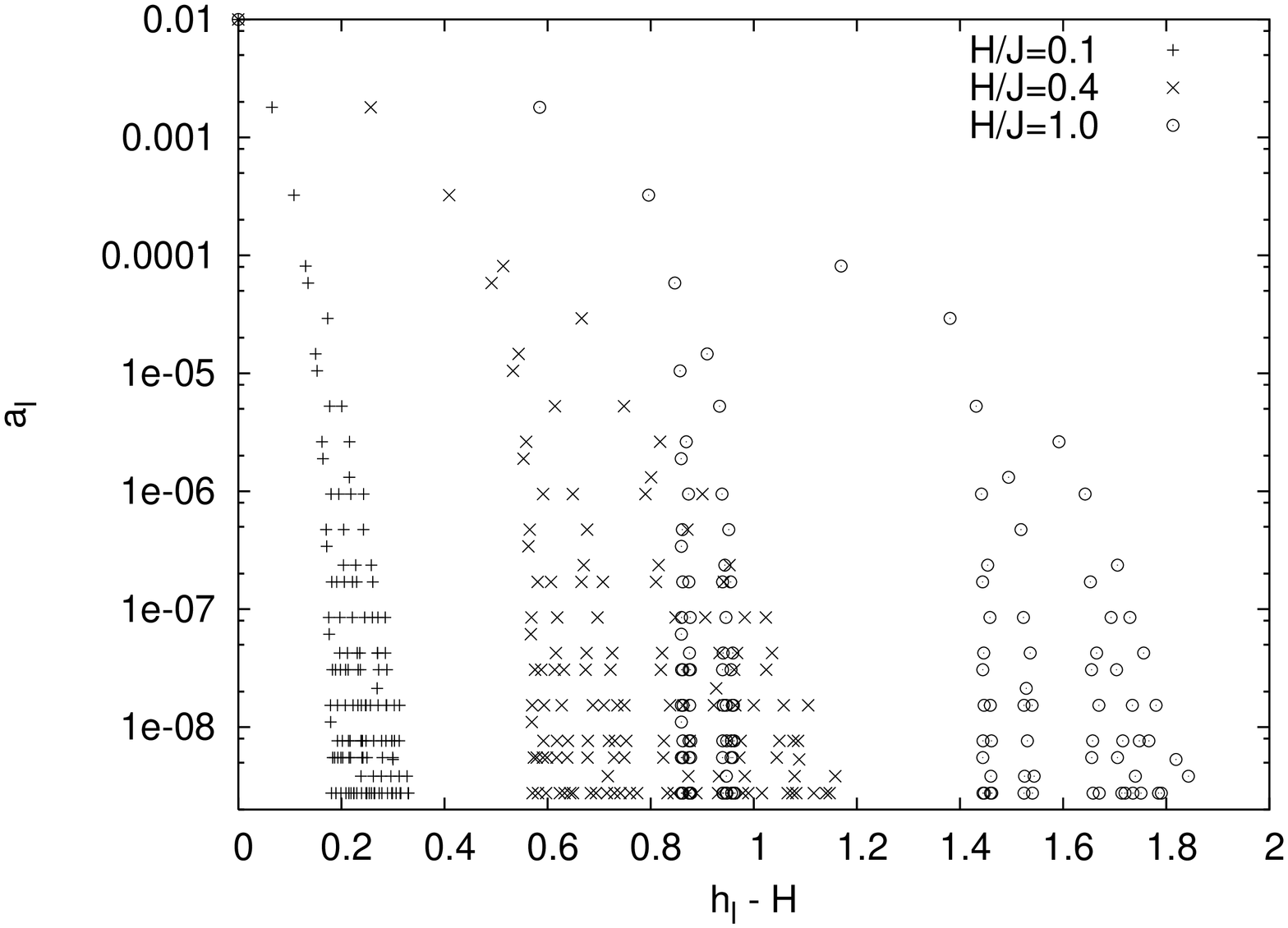}
\caption{\label{fig:discr}The coordinates and weights of delta peaks in the dicrete spectrum $s(h)$ for $k=2$, $\bar H=0$, $p_1=0.9$, $T/T_c=0.8$; $A_0=1/81$; the 132 heaviest levels are shown for $H/J=0.1$, $H/J=0.4$, and $H/J=1$. The weights for all the 3 cases are the same, only the positions of the delta peaks differ.}
\end{figure}

Let us note that the weights of delta functions depend only on parameter $p_1$, which is determined by the structure of the solid matrix. The positions of the delta functions depend also on other parameters ($T$, $H$, $\bar H$, and $J$). As we discussed in the previous section, the discrete spectrum corresponds to finite trees growing from a site. Each level in the spectrum corresponds to a tree of a certain size and form, and its weight is a probability to find such a tree in the system. Fig.~\ref{fig:discr} shows an example of the discrete spectrum for the case of a relatively sparse solid that blocks only a small fraction of sites. In that case the 20 `heaviest' levels carry more than 99.96\% of the total weight. As $p_1$ approaches the percolation threshold, the convergence of the series becomes progressively poorer. For example, the 200 heaviest levels in the case $k=2$, $p_1=0.6$, contain only 87\% of the discrete spectrum's weight. This naturally agrees with the fact that there appear a lot of large finite trees near the percolation point, and the number of forms of a tree quickly (combinatorially) grows with its size. One other interesting feature of the discrete spectrum is its bandlike structure, with  $a_l$ densely grouped around certain values, and $h_l$ subdividing into several branches as $H$ increases.

\begin{figure}
\includegraphics[height=0.7\textheight]{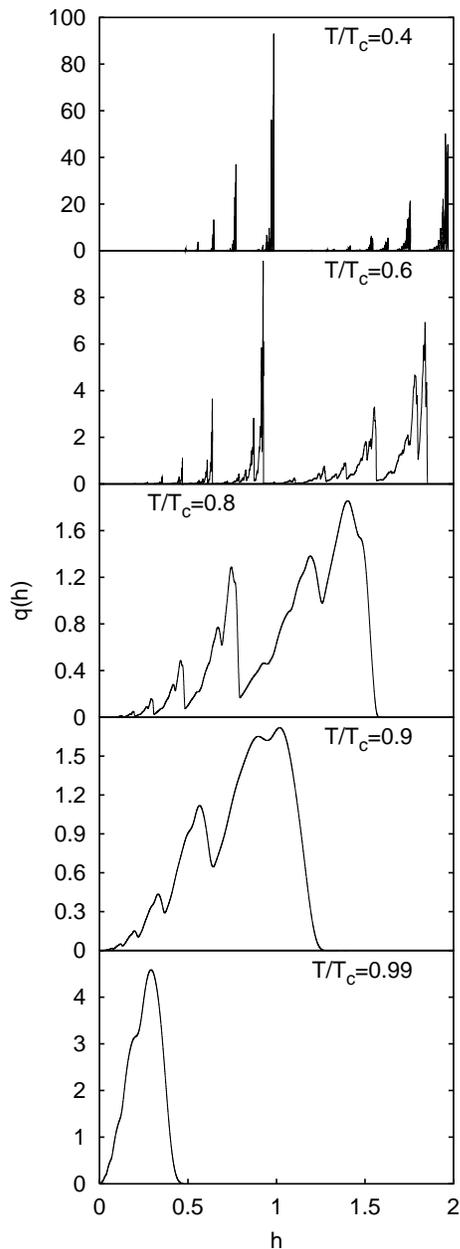}
\caption{\label{fig:q}The continuous part $q(h)$ of the effective field distribution for a weakly dilute Ising ferromagnet ($c_0=0.1$) with $H=0$, at different temperatures below the critical point.}
\end{figure}

\begin{figure}
\includegraphics[width=\columnwidth]{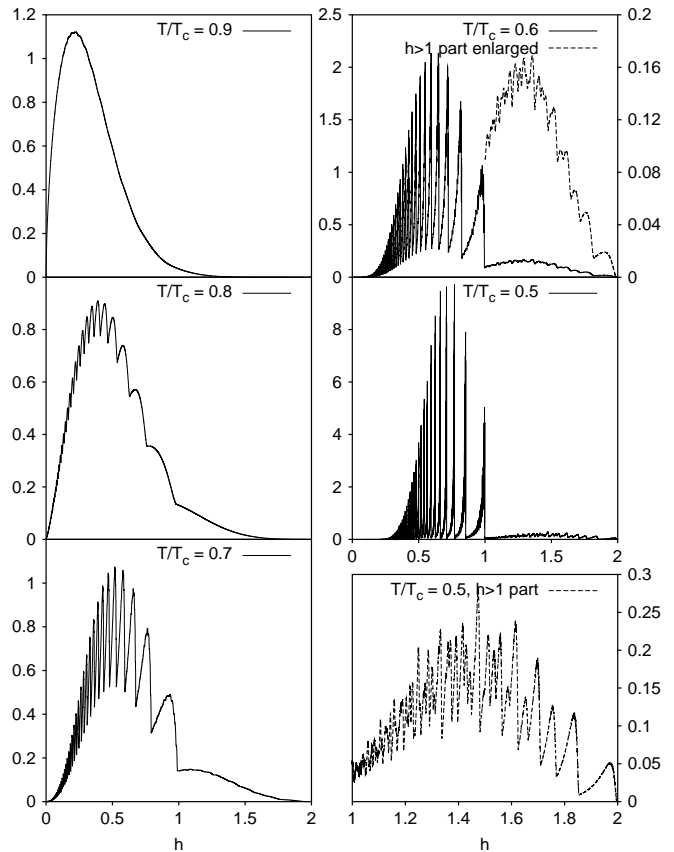}
\caption{\label{fig:q06}The continuous part $q(h)$ of the effective field distribution for a strongly dilute (but still percolated) Ising ferromagnet ($c_0=0.4$) in zero external field at different temperatures below the critical point. Some pictures contain a dashed line, which shows the $h>1$ part of $q(h)$ at an enlarged scale (plotted against the right $y$ axes).}
\end{figure}

Once $s(h)$ is calculated to a given high accuracy (defined by $\epsilon=(A_0-\sum_l a_l)/A_0$), we can proceed by solving equation \mref{contin}. A numerical algorithm must represent the function $q(h)$ as an $M$-vector $q(h_i)$, $h_i=h_\Min+i\Delta h$, $\Delta h=(h_\Max-h_\Min)/M$, $i=0\cdots M-1$, with $M$ sufficiently large. It is helpful to note that $q(h)$ is non-zero only on a finite interval $[h_\Min,h_\Max]$: this permits use of Fourier series instead of Fourier integrals. The lower and upper limits of this interval may be estimated as 
\begin{eqnarray}
&&h_\Min=H+\min(-kJ,(k-1)\bar H-J),
\nonumber\\
&&h_\Max=H+\max(kJ,(k-1)\bar H+J).
 \label{bounds}
\end{eqnarray}
These bounds are far from exact at high temperatures, but this does not make a  significant impact on the efficiency of the calculations \cite{footnote:improvedBounds}.

The numerical solution of Eq.~\mref{contin} is an iterative process. We start from some initial distribution (this may be a uniform distribution) and insert it into the \rhs of Eq.~\mref{contin}. The resulting \lhs is a new approximation to the solution. Each iteration produces the field distribution $q(h)$ one level deeper the tree. Therefore such iterations must converge provided the stable point distribution defined by Eq.~\mref{main} exists and numerical errors introduced by the discretization are small enough. 
Naive quadratures for Eq.~\mref{main} would result in a formula
\begin{eqnarray}
q(h_i)&=& 
 \sum_{j=1-M}^{M-1}\exp(i\xi_j(h_i-H))
 \times\nonumber\\&&\times
 [\{\tilde q(\xi_j)+\tilde s(\xi_j)\}^k-\tilde s^k(\xi_j)]
\label{contin2a},\\
\tilde q(\xi_j)&=&p_1\Delta h \sum_{l=0}^{M-1} q(h_l)\exp(-i\xi_j u(h_l))
\label{contin2b},\\
\xi_j&=&2\pi j/(h_\Max-h_\Min)
,\nonumber
\end{eqnarray}
which takes of order $M^3$ calculations at each iteration, because each of the indices $i$, $j$, and $l$ in \mref{contin2a} and \mref{contin2b} take $M$ essentially distinct values. This quadrature is both inefficient and incorrect. First, to implement the correct numerical algorythm, it is important to note that $\exp(-i\xi u(h'))$ is a quickly oscillating function of $h'$ when $\xi$ is large, and this requires a special quadrature when calculating the $h'$ integral in \mref{contin}. Second, the Fast Fourier Transform algorithm takes only $O(M\ln M)$ steps to perform the summation over $j$ in \mref{contin2a} for all $i$. Also, Press et al \cite{nr} describe an algorithm that permits simultaneous calculation of $M$ integrals over $h'$ in \mref{contin} using again only $O(M\ln M)$ operations. Therefore, with advanced algorithms, each iteration costs $O(M\ln M)$ arithmetic operations. This is a dramatic gain with respect to the quadratures estimate of $O(M^3)$, because as seen below, $M$ is typically many thousand.

\begin{figure}
\includegraphics[angle=-90,width=\columnwidth]{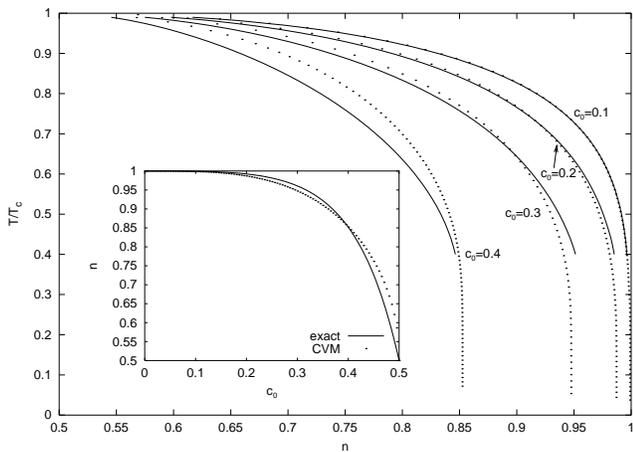}
\caption{\label{fig:dilBinodal}The high-density branch of the GCE binodal ($n_+$) for the model with $\bar H=0$ at different matrix densities $c_0$. The low density branch is symmetric: $n_-=1-n_+$. The percolation concentration for the lattice used ($k=2$) is $c_0=1/2$. The inset shows the dependence of $n_+$ at zero temperature on the matrix density. The dotted lines represent the corresponding results of the cluster variation method \cite{TCA}.}
\end{figure}

The numerical results we report here are limited to coordination number 3 ($k=2$), and a completely random matrix, when there is no correlation between quenched variables even at the nearest neighbor sites: hence $w_{ab}=c_ac_b$ and $p_a=c_a$. Since we are studying the behavior of a fluid in a porous media, we can set aside very dense matrices with $p_1<1/k$; these have no infinite connected volume accessible to the fluid, so that $q(h)=0$, and thermodynamic potential \mref{Ffluid} equals zero. 

First we shall consider the simpler case of a diluted Ising ferromagnet ($\bar H=0$), which corresponds to fluid in the porous matrix with a moderate fluid-matrix attraction ($K=I/2$). In this case the exact critical temperature is known from Eq.~\mref{Tc}; when $H=0$ and $T>T_c$, $Q(h)=\delta(h)$ and $m=0$ ($n=1/2$), while at lower temperatures there are two symmetric non-trivial solutions $Q_+(h)=Q_-(-h)$, with $Q_+(h)=A_0\delta(h)+q_+(h)$, $A_0=(c_0/c_1)^2$. The temperature dependence of $q_+(h)$ is shown in Fig.~\ref{fig:q}. One can see that the $q_+(h)$ line repeats itself, but not in a trivial fashion. The continuous field distribution acquires even more structure for denser matrices (Fig.~\ref{fig:q06}). At low temperatures $q_+(h)$ becomes a set of increasingly sharp peaks, and this greatly increases the errors introduced by the discretization in the numerical algorithm. Therefore this algorithm does not work at very low temperatures. Nevertheless, using extremely fine discretization (we used $M$ up to $2^{17}=131072$), we covered an interesting range of temperatures. With the algorithm just described we were able to reach $T/T_c=0.4$. In the magnetic interpretation of the model \mref{Ham} the solutions $Q_+(h)$ and $Q_-(h)$ correspond to exactly opposite magnetizations ($m_+=-m_-$), which reflects the fact that direction of the spontaneous (zero-field) magnetization is not predetermined. In the fluid interpretation these two solutions form the two branches of the binodal (GCE:~$n_+=(m_++1)/2$, $n_-=1-n_+$). Fig.~\ref{fig:dilBinodal} shows the resulting GCE binodals at different densities of the matrix.  According to Eq.~\mref{dn}, the `infinite cluster' binodals for this particular case ($\bar H=0$) have the same size and shape as GCE binodals, they are only shifted to lower densities by $\Delta n=x/2$, $\Delta n=6.8 \cdot 10^{-4}$, $7.8\cdot 10^{-3}$, 0.039, 0.15 for $c_0=0.1$, 0.2, 0.3, and 0.4, respectively. In the general case ($\bar H\ne0$), the density shift between the GCE and IC binodals depends on temperature.

The binodal density (or zero field magnetization, in the Ising language) at zero temperature can be found from the fact that in this case all the spins in the infinite cluster are aligned, whereas finite clusters do not contribute to the spontaneous magnetization, and the magnetization (per magnetic site) equals plus or minus the fraction of magnetic sites belonging to the infinite cluster: $m_+(T=0)=1-x$. This gives the resulting half-width of the binodal, which is depicted in the inset of Fig.~\ref{fig:dilBinodal} as a function of matrix density $c_0$. One can see that the cluster approximation produces the binodals that are flatter at the top and have an incorrect width. For a lattice with only sparsely distributed solid sites, the cluster approximation results are quite accurate.

\begin{figure}
\includegraphics[width=\columnwidth]{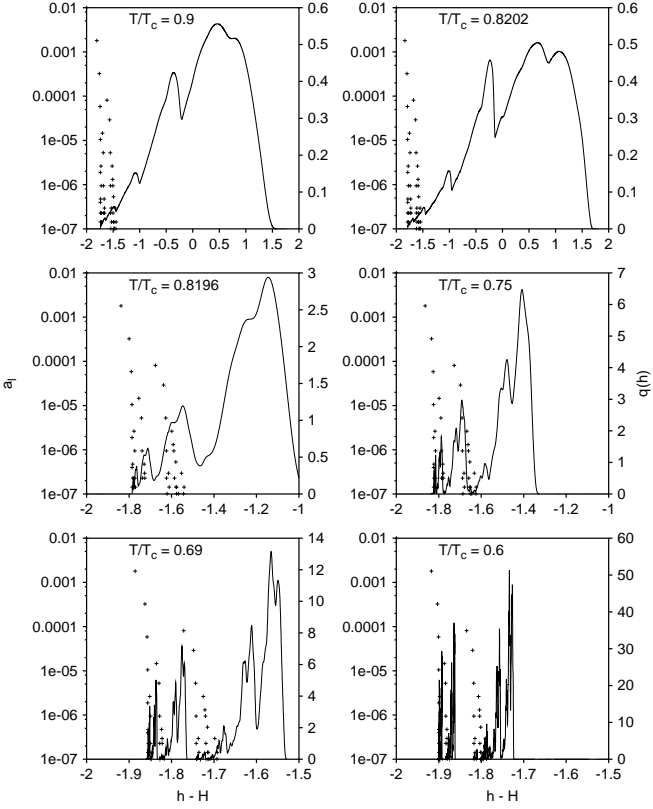}
\caption{\label{fig:qhs1}The effective field distribution $Q(h)$ for a fluid in a sparse hard-core matrix ($\bar H=-J$, $c_0=0.1$) at $H=0.21J$ at different temperatures. The crosses show weights $a_l$ and locations $h_l$ of leading delta functions in the discrete part of spectrum $s(h)=\sum_l a_l\delta(h-h_l)$, plotted against the left $y$ axes. The lines depict the continuous part of the distribution $q(h)$ (against the right $y$ axes). The $x$ axis is $h-H$. }
\end{figure}

\begin{figure}
\includegraphics[angle=-90,width=\columnwidth]{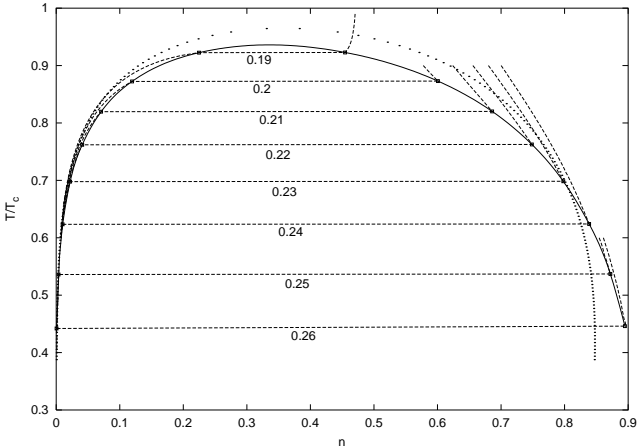}
\caption{\label{fig:HSBinodal}The temperature dependence of the equilibrium fluid density at different values of $H$ for the model with $\bar H=-J$ at matrix density $c_0=0.1$. The numbers at the dashed lines are the values of the external field $H$. The lines jump over the phase separation region, delimited by the solid line. The binodal within the cluster approximation \cite{unpub} is depicted with a dotted line.}
\end{figure}

Now, let us consider instead the situation where there is no matrix-fluid attraction (hard core matrix): $K=0$, $\bar H=-J$. Fig.~\ref{fig:qhs1} shows how the field distribution at constant chemical potential (fixed $H$) now changes with temperature. In the case depicted there is a competition between the surface field $\bar H=-J<0$ and the bulk field $H=0.21J>0$. At high temperatures the effective field prefers the sign of the bulk field and is mainly positive. At $T/T_c<0.82$, an alternative solution for $q(h)$, which favors negative fields, corresponds to the global minimum of the thermodynamic potential and becomes preferable. At $T/T_c=0.82$ the two solutions yield the same value of thermodynamic potential at different fluid densities, and correspond to the two branches of the binodal. It should be noted that $T_c$ is defined as given by Eq.~\mref{Tc} and is not at (but is close to) the critical temperature for this matrix. Fig.~\ref{fig:HSBinodal} shows the temperature dependence of the fluid density at different values of $H$. The jumps correspond to coexistent densities at the binodal. One can see that in this case the binodal is not symmetric. The results depicted in Fig.~\ref{fig:HSBinodal} correspond to the `grand canonical fluid' described by Eqs.~\mref{Fav}, \mref{nGC}. The fraction of the free sites that belong to finite closed cavities ($x=(c_0/c_1)^3$) is as small as $0.14\%
$ for this relatively sparse matrix, therefore the binodal of the `infinite cluster fluid', which excludes their contribution, is indistinguishable on the plot from that depicted by the whole line in Fig.~\ref{fig:HSBinodal}. Let us also note that due to a special symmetry of the model \ref{IsHam} \cite{footnote:symmetry}, the binodal of the symmetrical case $\bar H=J$ ($K=2J$, attractive matrix) coinsides with that in Fig.~\ref{fig:HSBinodal} flipped with respect to $n=1/2$ line.

\begin{figure}
\includegraphics[angle=-90,width=\columnwidth]{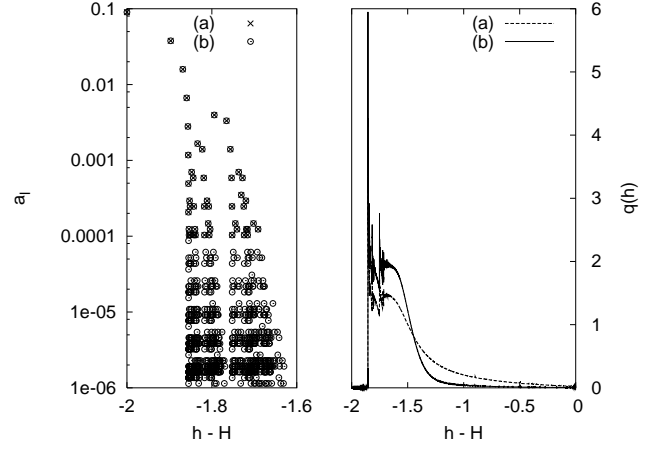}
\caption{\label{fig:incompleteS}The field distribution for the model with $\bar H=-J$ at matrix density $c_0=0.3$, $H=0.61$, and $T/T_c=0.64$ calculated with truncated discrete spectrum $s(h)$. One of the cases depicted (a) corresponds to 48 leading delta-functions taken into account in $s(h)$ (all the terms with weights $a_l>10^{-4}$). In the other case (b) there are 804 terms in $s(h)$ (the terms with weights $a_l>10^{-6}$)}
\end{figure}

Next, we would like to consider denser matrices, and here the poor convergence of $s(h)$ leads to a problem. We did not have this problem in the case of dilute magnet ($\bar H=0$), because its entire binodal, in view of spin-flip symmetry of that special case, corresponded to $H=0$, when the discrete spectrum consisted of only one term: $s(h)=A_0\delta(h)$, which could be exactly calculated for the matrix of any density (Eq.~\mref{A0}). For the hard core matrix ($\bar H=-J$) this no longer takes place. For $c_0=0.3$ the 48 leading delta-functions of $s(h)$ contain 95.5\% of its weight, 804 terms contain 98.6\%. If we simply ignore the residual weight and solve Eq.~\mref{contin} with an incomplete $s(h)$, the resulting $q(h)$ strongly depends on the number of terms taken into account in $s(h)$ (see Fig.~\ref{fig:incompleteS}).

\begin{figure}
\includegraphics[angle=-90,width=\columnwidth]{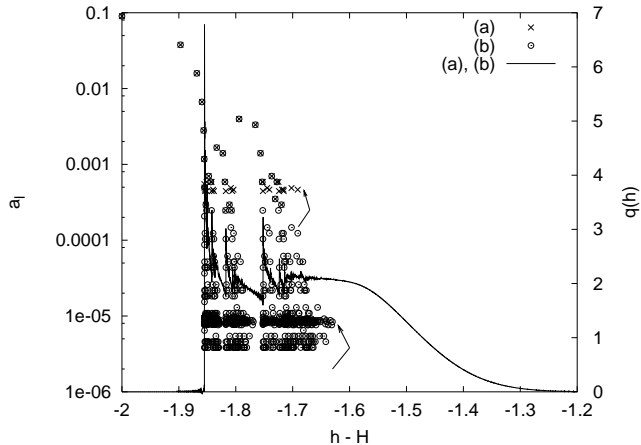}
\caption{\label{fig:adjustedS}The same as Fig.~\ref{fig:incompleteS}, but calculated with an adjusted discrete spectrum $s(h)$, Eq.~\mref{adjustedS}. The arrows point to the groups of $a_l$'s raised by $\Delta$. The resulting continuous spectra corresponding to (a) 48 and (b) 804 terms in $s(h)$ are indistinguishable. }
\end{figure}

To overcome the problem, we use the following trick. First we calculate the leading $2l$ terms of $s(h)$. They correspond to $2l$ relatively small finite trees, because the larger trees have smaller weight. Then we determine the $l$ lightest terms among them and add $\Delta=(A_0-\sum_{i=1}^{2l}a_i)/l$ to their weights. The resulting discrete distribution 
\begin{equation}
s(h)=\sum_{i=1}^l a_i\delta(h-h_i)+\sum_{i=l+1}^{2l} (a_i+\Delta)\delta(h-h_i)
\label{adjustedS}
,
\end{equation}
has the correct total weight. This way we distribute the missing weight of large trees among the largest of those taken into account. From the physical point of view this means that we replace the `very large' branches by the `large' ones and in this way forbid `very wide' pores. The latter implies a certain correlation in the matrix and is always an approximation, since when deriving equation \mref{contin} we permitted only the nearest neighbor matrix correlations. Anyway, the results must become accurate in the limit $l\to\infty$, and we find that the method converges already at small $l$ (Fig.~\ref{fig:adjustedS}). A remarkable new feature of $q(h)$ in Fig.~\ref{fig:adjustedS} is the presence of steps, accompanied by sharp peaks, especially visible at $h-H=-1.85$ or $-1.75$. It is possible that these peaks are not accurate (see Appendix~\ref{app:B}), they could be numerical artefacts arising in places where $q(h)$ in reality has only a step \cite{footnote:terminology}. Note anyway that these peaks are not delta functions: their heights remain almost unchanged when we increase resolution from $M=2^{14}$ to $M=2^{15}$, whereas if they were delta peaks, their heights would then double \cite{footnote:delta}.

\begin{figure}
\includegraphics[width=\columnwidth]{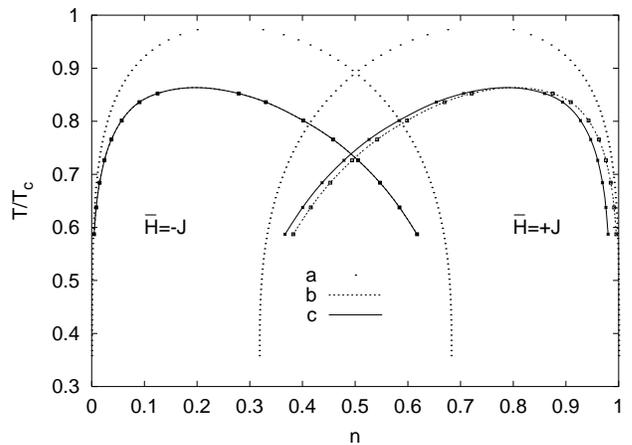}
\caption{\label{fig:HS02Binodal}The binodals for the models with $\bar H=-J$ and $\bar H=J$ at matrix density $c_0=0.2$. The bold lines (c) represent the `infinite cluster' binodals. The thin dashed lines (b) are the grand canonical ensemble binodals (in the $\bar H=-J$ case this line is hidden by the overlapping (c) line). The binodals within the cluster approximation \cite{unpub} are depicted with a dotted line (a). The pairs of (a) and (b) binodals are symmetric with respect to $n=1/2$. The pair of (c) lines is symmetric with respect to $n=(1-x)/2=0.492$. The small squares on (b) and (c) lines are the actual data points through which the interpolating lines were drawn.}
\end{figure}

\begin{figure}
\includegraphics[width=\columnwidth]{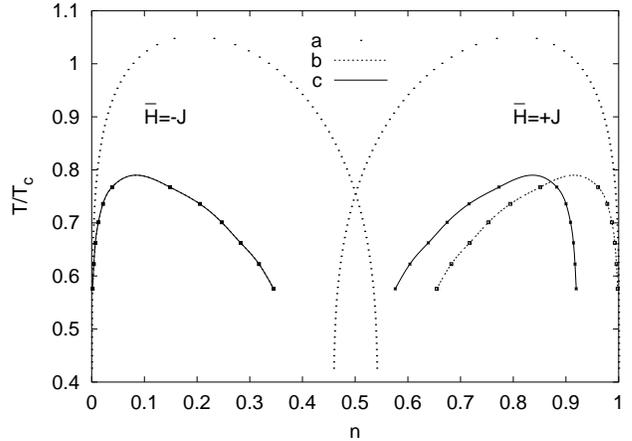}
\caption{\label{fig:HS03Binodal}The same as Fig.~\ref{fig:HS02Binodal}, at matrix density $c_0=0.3$. The pair of (c) binodals is symmetric with respect to $n=(1-x)/2=0.461$.}
\end{figure}

Having overcome this convergence problem with $s(h)$, we can calculate phase diagrams for denser matrices (Figs.~\ref{fig:HS02Binodal}-\ref{fig:HS03Binodal}). For these denser matrices, unlike the case $c_0=0.1$, the difference between the grand canonical ensemble binodals and the infinite cluster ones becomes clearly visible, although the fraction of finite closed cavities is still not large ($x(c_0=0.2)=1.6\%
$, $x(c_0=0.3)=7.9\%
$). This fraction grows to 100\% at the percolation point (which happens at $c_0=0.5$ for this lattice), thus making a big difference near percolation.

One can see that GCE and IC binodals for $\bar H=-J$ are indiscernable, which indicates that for this case closed pores in GCE are almost empty, and their contribution is negligible \cite{footnote:negligible}. On the contrary, for  $\bar H=J$, they are almost saturated with fluid, and $\Delta n$ is close to the fraction $x$ of closed pores in the matrix. This situation should also hold for strongly attractive or repulsive matrices: $|\bar H|>1$. For dense enough matrices with smaller $|\bar H|$, $\Delta n$ will be nonzero for both negative and positive $\bar H$ and will depend on temperature.

In the above results for our numerical treatment of the exact integral equations, deviations from the cluster approximation are clearly visible in the phase diagrams (Figs.~\ref{fig:dilBinodal}, \ref{fig:HSBinodal}), and become very large in some cases (Figs.~\ref{fig:HS02Binodal}, \ref{fig:HS03Binodal}). Note that the cluster approximation permits no simple way round the failure of the grand canonical ensemble to prevent fluid from entering closed pores.

It is interesting to note that some other related approaches \cite{Kierlik9799,Kierlik98} predict an unusual double-hump binodal of the fluid in moderately dense strongly attractive or repulsive matrices (in terms of the current model, this means large $|\bar H|$). Kierlick et al \cite{Kierlik9799} considered an off-lattice molecular model of fluid in a quenched disordered configuration of spheres on the basis of a replica symmetric Ornstein-Zernike equation. They studied a sequence of approximations, the first of which corresponded to the mean spherical approximation (MSA), and predicted a binodal of the usual form. The higher approximation yielded a binodal with two critical points or even two disconnected binodals. Their highest order approximation again gave only one critical point, but predicted a shoulder on one of the binodal's sides. The MSA applied to the model \mref{Ham} on the five-dimensional fcc or bcc lattice \cite{Kierlik98} again predicted a double-humped binodal. A problem identified for MSA was that for the 3-dimensional lattices it predicted a double-humped binodal even for the bulk fluid without any porous matrix. Therefore the authors of \cite{Kierlik9799,Kierlik98} could not conclude definitely about the existence of two critical points for these models, despite the similarity of their results to the double-humped binodals seen in Monte-Carlo simulations on similar models \cite{Page96,Alvarez99}. Our theory does not give any double-humped binodal, at least in the range of parameters we have studied.

\section{\label{sec:conclusion}Conclusions}
We considered the Bethe lattice model of fluid in a porous medium. The recursive character of the Bethe lattice permits an exact treatment, whose key ingredient is an integral equation for the effective field distribution. Solutions of this equation consist of a sum of discrete and continuous spectra, and these spectra have distinct physical interpretations. The discrete spectrum comes from disconnected finite pore spaces, whereas the continuous spectrum is a contribution of the infinite pore space which in reality is the only one accessible to fluid. The continuous spectrum develops more and more structure at low temperature, which means that a numerical solution for it becomes impractical below a certain temperature. However, the physical results found for temperatures above this threshold are both consistent and reasonable. Despite use of a Bethe lattice they differ significantly from results calculated using the cluster (or Bethe) approximation, which cannot handle the complexity of the field distributions that we find.

The dichotomy between the two types of pore space mentioned above is not exclusive to the Bethe lattice, but universal. Any microscopic model of fluid in a random porous media that uses the grand canonical ensemble will include contributions of the finite cavities, unless this is carefully subtracted off, as we managed to do for the Bethe lattice. In the grand canonical ensemble, these cavities are in equilibrium with the external bulk fluid, but in real-world experiments they are inaccessible and do not respond to changes of chemical potential. This marks an important distinction between models of fluids in porous media and disordered magnetic model to which they are equivalent in the grand canonical ensemble; for magnetism, finite clusters do contribute to the free energy and it is right to include them.

\begin{acknowledgments}
R.~O.~S. thanks NATO and the Royal Society for funding his visit to the University of Edinburgh. R.~O.~S. also thanks School of Physics of the University of Edinburgh for financial support of this visit. T.~G.~S and R.~O.~S. thank the Condensed Matter Group of the University of Edinburgh for hospitality.
\end{acknowledgments}

\appendix

\section{\label{app:A}Solution of the integral equation for the field distribution for the one-dimensional chain}

In the one-dimensional chain $k=1$, and Eq.~\mref{main} can be solved analytically. Performing first the integration with respect to $\xi$ in the \rhs of Eq.~\mref{main}, one gets
\begin{widetext}
\begin{eqnarray}
Q(h)&=&p_0\delta(h-H-\bar H)+p_1\int\di h'Q(h')\delta(h-H-u(h'))
\nonumber\\
&=&p_0\delta(h-H-\bar H)+p_1\dot v(h-H)Q(v(h-H))
,\label{a:Q1}
\end{eqnarray}
\end{widetext}
where $v(h)$ is the function inverse to $u(h)$, that is $v(u(h))\equiv h$, and $\dot v(h)$ is its derivative. The above equation says that $Q(h)$ equals $p_0\delta(h-H-\bar H)$ for $h$ in the close vicinity of $H+\bar H$, otherwise one has to look at the value of $Q$ at $v(h-H)$ and multiply it by $p_1\dot v(h-H)$. This observation leads to a solution of the form $Q(h)=\sum_la_l\delta(h-h_l)$. The leading delta function is given explicitly in the \rhs of Eq.~\mref{a:Q1}, and positions of the further delta functions are given by relations
\begin{equation}
v(h_2-H)=h_1; \; v(h_3-H)=h_2; \; \cdots
\label{a:rec1}
\end{equation}
This means, in particular, that in the vicinity of $h_2$ 
\begin{eqnarray}
Q(h)&=&p_1\dot v(h-H)Q(v(h-H))
\nonumber\\
&=&p_1\dot v(h-H)p_0\delta(v(h-H)-h_1)
\nonumber\\
&=&p_1p_0\delta(h-h_2),
\end{eqnarray}
and therefore $a_2=p_1p_0$. In general, one can find that $a_l=p_1a_{l-1}$. Finally, we can write the solution in the form
\begin{equation}
Q(h)=p_0\sum_{l=1}^\infty p_1^{l-1}\delta(h-h_l)
\end{equation}
where $h_l$ are given by the recursion that follows from Eq.~\mref{a:rec1},
\begin{equation}
h_l=u(h_{l-1})+H;\; l\ge2 ;\; h_1=H+\bar H.
\end{equation}

\section{\label{app:B}On the nature of spikes in the continuous distribution $q(h)$}
\begin{figure}
\includegraphics[width=0.8\columnwidth]{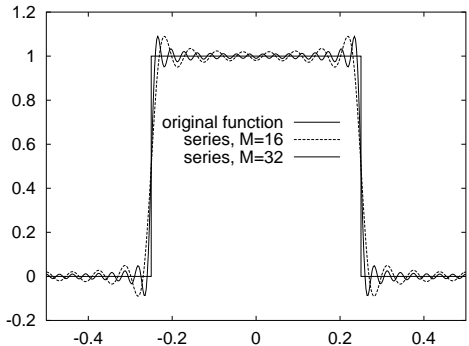}
\caption{\label{fig:FourierExpansion}A function with discontinuities (periodic on $[-0.5;0.5]$) and its Fourier expansion, truncated at $M=16$ and $M=32$.}
\end{figure}

In our numerical calculation we represent $q(h)$ as an $M$-vector and use truncated Fourier series. This is guarantied to work well only for smooth functions. For example, truncated Fourier series result in peaks near discontinuities (Fig.~\ref{fig:FourierExpansion}). These are called `Gibbs ears'. Note that Gibbs ears have a height that remains fixed, while their width narrows to zero as the resolution improves. There is accordingly no area beneath a Gibbs ear, unlike a delta function. The spikes in Fig.~\ref{fig:adjustedS} do not look exactly like Gibbs ears: Gibbs ears appear in pairs, and the positive ear has a negative counterpart, whereas spikes in Fig.~\ref{fig:adjustedS} are asymmetric, with negative spikes much smaller and almost absent at the fine discretization ($M=2^{16}$) that we used to produce the plot. It is still possible that equation/recursion \mref{contin} somehow creates a type of positive-only Gibbs ear effect, but this does not seem likely.

\begin{thebibliography}{99}
\bibitem{footnote:limit}	Of course, we are interested in the thermodynamic limit: $V\to\infty$.
\bibitem{footnote:case}This corresponds to the case when the matrix-fluid attraction equals one-half of the fluid-fluid attraction: $K=I/2$. In this case the model is spin flip symmetric at $H=0$, which results in the symmetric binodal with critical point at $n=1/2$.
\bibitem{footnote:precise}To be precise, this fraction equals $c_1p_0^z$.
\bibitem{footnote:correspondence}There is a one to one correspondence between $s(h)$ and $\tilde s(\xi)$ through
Eq.~\mref{discr}.
\bibitem{footnote:delta}This is also true for the peaks in Fig.~\ref{fig:q} ($T/T_c=0.4$). Their height only slightly increases upon increasing the resolution from $2^{16}$ to  $2^{17}$.
\bibitem{footnote:negligible} It means only that \textit{the contribution to $n$} is very small in this case. If we consider the density per \textit{accessible} volume $n'=n/(1-x)$, where inaccessible fraction $x$ is defined by Eq.~\mref{xx} for IC fluid, and $x=0$ for GCE (since closed pores are accessible in GCE), the difference $\Delta n'=n'_{\rm GCE}-n'_{\rm IC}$ is noticeable for both $\bar H=J$ and $\bar H=-J$ cases. In fact, $\Delta n'_{\bar H=J}=-\Delta n'_{\bar H=-J}$
\end{thebibliography}
\end{document}